\documentclass[acmlarge,screen, nonacm]{acmart}

\AtBeginDocument{%
  \providecommand\BibTeX{{%
    \normalfont B\kern-0.5em{\scshape i\kern-0.25em b}\kern-0.8em\TeX}}}

\usepackage{tikz}
\usepackage{amsmath}

\usepackage{tabularx}
\usepackage{booktabs}
\usepackage{wasysym}
\usepackage{forloop}
\usepackage{ifthen}
\usepackage{enumitem}
\usepackage{url}
\PassOptionsToPackage{hyphens}{url}
\usepackage{hyperref}
\usepackage{caption}
\usepackage{xspace}
\usepackage{ulem}
\usepackage{makecell}

\setitemize{noitemsep,topsep=0pt,parsep=0pt,partopsep=0pt}

\usetikzlibrary{shadings}
\usetikzlibrary{patterns}
\definecolor{NiceGreen}{RGB}{27,158,119}
\definecolor{NiceRed}{RGB}{217,95,2}

\newcommand{\systemname}{SkillBot\xspace}

\newif\ifrev
\ifrev
    \newcommand{\yuan}[1]{\textcolor{red}{[Yuan: #1]}}
    \newcommand{\danny}[1]{\textcolor{orange}{[Danny: #1]}}
    \newcommand{\dannyedit}[1]{\textcolor{orange}{#1}}
    \newcommand{\tu}[1]{\textcolor{blue}{[Tu: #1]}}
    \newcommand{\noah}[1]{\textcolor{purple}{[Noah: #1]}}
\else
    \newcommand{\yuan}[1]{}
    \newcommand{\danny}[1]{}
    \newcommand{\dannyedit}[1]{}
    \newcommand{\tu}[1]{}
    \newcommand{\noah}[1]{}
\fi

\newif\ifdiff
\newcommand{\revise}[1]{%
\ifdiff
\textcolor{blue}{#1}%
\else
#1%
\fi
}

\newcommand{\Qq}[1]{\textbf{#1}}

\newcounter{qr}

\newcommand{\Qline}[1]{\noindent\rule{#1}{0.6pt}}

\newcounter{ql}

\newenvironment{Qlist}{%

\begin{itemize}[topsep=0.1em]
}{%
\end{itemize}
}

\newlength{\qt}

\newcounter{itemnummer}
\newcommand{\Qitem}[2][]{
\ifthenelse{\equal{#1}{}}{\stepcounter{itemnummer}}{}
\ifthenelse{\equal{#1}{a}}{\stepcounter{itemnummer}}{}
\begin{enumerate}[topsep=2pt,leftmargin=2.8em]
\item[\textbf{\arabic{itemnummer}#1.}] #2
\end{enumerate}
}

\definecolor{bgodd}{rgb}{0.8,0.8,0.8}
\definecolor{bgeven}{rgb}{0.9,0.9,0.9}
\newcounter{itemoddeven}
\newlength{\gb}
\newcommand{\QItem}[2][]{
\setlength{\gb}{\linewidth}
\addtolength{\gb}{-5.25pt}
\ifthenelse{\equal{\value{itemoddeven}}{0}}{%
\noindent\colorbox{bgeven}{\hskip-3pt\begin{minipage}{\gb}\Qitem[#1]{#2}\end{minipage}}%
\stepcounter{itemoddeven}%
}{%
\noindent\colorbox{bgodd}{\hskip-3pt\begin{minipage}{\gb}\Qitem[#1]{#2}\end{minipage}}%
\setcounter{itemoddeven}{0}%
}
}

\begin{document}

\title{SkillBot: Identifying Risky Content for Children in Alexa Skills}

\author{Tu Le}
\affiliation{%
  \institution{University of Virginia}
  \country{USA}
}
\author{Danny Yuxing Huang}
\affiliation{%
  \institution{New York University}
  \country{USA}
}
\author{Noah Apthorpe}
\affiliation{%
  \institution{Colgate University}
  \country{USA}
}
\author{Yuan Tian}
\affiliation{%
  \institution{University of California, Los Angeles}
  \country{USA}
}

\renewcommand{\shortauthors}{Le et al.}

\begin{abstract}
Many households include children who use voice personal assistants (VPA) such as Amazon Alexa. Children benefit from the rich functionalities of VPAs and third-party apps but are also exposed to new risks in the VPA ecosystem. In this paper, we first investigate ``risky'' child-directed voice apps that contain inappropriate content or ask for personal information through voice interactions. We build SkillBot --- a natural language processing (NLP)-based system to automatically interact with VPA apps and analyze the resulting conversations.
We find 28 risky child-directed apps and maintain a growing dataset of 31,966 non-overlapping app behaviors collected from 3,434 Alexa apps. Our findings suggest that although child-directed VPA apps are subject to stricter policy requirements and more intensive vetting, children remain vulnerable to inappropriate content and privacy violations. We then conduct a user study showing that parents are concerned about the identified risky apps. Many parents do not believe that these apps are available and designed for families/kids, although these apps are actually published in Amazon's ``Kids'' product category. We also find that parents often neglect basic precautions such as enabling parental controls on Alexa devices. Finally, we identify a novel risk in the VPA ecosystem: confounding utterances, or voice commands shared by multiple apps that may cause a user to interact with a different app than intended. We identify 4,487 confounding utterances, including 581 shared by child-directed and non-child-directed apps. We find that 27\% of these confounding utterances prioritize invoking a non-child-directed app over a child-directed app. This indicates that children are at real risk of accidentally invoking non-child-directed apps due to confounding utterances.
\end{abstract}



\setcopyright{acmcopyright}
\acmJournal{TOIT}

\maketitle

\section{Introduction}
\label{section:introduction}

The rapid development of Internet of things (IoT) technology has aligned with growing popularity of voice personal assistant (VPA) services, such as Amazon Alexa and Google Home. 
In addition to the first-party features provided by these products, 
VPA service providers have also
developed platforms that allow third-party developers to build and publish their own voice apps---hereafter referred to as ``skills''.

\paragraph{\textbf{Risks to Children from VPAs.}}
Researchers have found that 91\% of children between ages 4 and 11 in the U.S. have access to VPAs, 26\% of children are exposed to a VPA between 2 and 4 hours a week, and 20\% talk to VPA devices more than 5 hours a week~\cite{collins}. 
The lack of robust authentication on commercial VPAs makes it challenging to regulate children's use of VPA skills~\cite{yuan2018echoattack}, especially as anyone in the physical vicinity of a VPA can interact with the device.
Additionally, parental control modes provided by VPAs (e.g., Amazon FreeTime and Google Family App) often place a burden on parents during setup and receive complaints from parents due to their limitations~\cite{redditfreetime, amazonforumfreetime, redditfreetimeunlimited}.

Legal and industry efforts have tried to protect children using VPAs; however, their effectiveness is unclear. 
In the U.S., the 1998 Children's Online Privacy Protection Act (COPPA) places information collected online from children under the age of 13 under parental control~\cite{ftccoppafaq}, but widespread COPPA violations have been shown in the mobile application market~\cite{reyes2018won}, and compliance in the VPA space is far from guaranteed. 
Other work has studied parents' concerns about their children's Internet use~\cite{sorbring2014parents}, including reports that more than 80\% of U.S.~parents are concerned about their children having access to sexual content, making friends with strangers, and exposing personal information~\cite{rosen2008association}. 

In this paper, we consider two types of threats posed by VPAs to children: (1) risky skills and (2) confounding utterances. We define ``risky'' skills as skills that contain inappropriate content for children or ask for personal information through voice interactions. An example is the ``My Burns'' skill in Amazon's Kids category that says ``You're so ugly you'd scare the crap out of the toilet.''
Another example is a skill named ``Shape Game'' that asks for age information: ``Awesome! Before we start however; I’m curious...how old are you?''
We define ``confounding utterances'' as voice commands shared among two or more skills that could cause a user to unintentionally invoke a different skill than intended.

Anyone can create a developer account and publish their skills to the Alexa Skills store for free. Additionally, the Alexa Skills Kit developer console~\cite{alexaskillskit} makes it easy to develop, validate, and publish skills even with limited experience. Therefore, risky skills and confounding utterances could be the result of intentionally malicious developers or benign/inexperienced developers unaware of potential risks.

\paragraph{\textbf{Challenges to Automated Skill Analysis.}}
It is challenging to systematically evaluate VPA skills to identify those that contain risky content or confounding utterances.
Substantial prior research has focused on COPPA compliance~\cite{reyes2017our, Reyes2018WontST} and large-scale measurements~\cite{razaghpanah2018apps} in the mobile application domain.
For example, Reyes et al.~\cite{reyes2017our, Reyes2018WontST} showed that many mobile apps on the Google Play Store and Apple App Store appeared to violate COPPA and infringe on privacy rights.
Razaghpanah et al.~\cite{razaghpanah2018apps} similarly identified and characterized organizations associated with mobile advertising and user tracking. 

However, little large-scale automated evaluation has been performed in the VPA domain. 
Unlike mobile apps that users download and install on their smartphones, VPA skills are essentially web apps hosted on VPA service providers' cloud servers. 
Therefore, VPA skills do not provide binary files or installation packages for download.
This makes empirical analysis of VPA skills challenging, as neither the executable files nor source code of VPA skills are available to researchers.
Most existing techniques and frameworks for automated analysis of mobile applications employ static or dynamic analysis of binary or source code and cannot be applied to VPA skills. 

VPA skills' natural language processing modules and key function logic are hosted in the cloud as a black box.  
VPA skill voice interactions are built following a template defined by the third-party developer, which is also unavailable to researchers. 
To automatically detect risky content, we need to generate testing inputs that trigger this content through sequential interactions. 
A further challenge is that risky content does not always occur during a user's first interaction with a skill; human users often need to have back-and-forth conversations with skills to discover risky content. 
Automating this process requires developing a tool that can generate 
valid voice inputs and dynamic follow-up responses that will cause the skill to reveal risky content. This is different from existing chatbot techniques~\cite{harkous2016pribots}, as the goal is not to generate inputs that sound natural to a human. 
Instead automated skill analysis requires generating inputs that explore the space of skill behaviors as thoroughly as possible.

\paragraph{\textbf{Research Questions.}} The risks and challenges discussed above mean that protecting children in the era of VPAs raises several pressing questions: 
\begin{itemize}
    \item \textbf{RQ0.} Can we automate the analysis of VPA skills to identify risky content for children without requiring manual voice interactions?
    \item \textbf{RQ1.} Are VPA skills specifically targeted to children that claim to follow additional content requirements -- hereafter referred to as ``kid skills'' -- actually safe for child users?
    \item \textbf{RQ2.} What are parents' attitudes and awareness of the risks posed by VPAs to children?
    \item \textbf{RQ3.} How likely is it for children to be exposed to risky skills through ``confounding utterances''---voice commands shared by multiple skills that could cause a child to accidentally invoke or interact with a different skill than intended. 
\end{itemize}  
In this paper, we address the challenges to analyzing skills at scale. We design, implement, and perform a systematic automated analysis of the Amazon Alexa VPA skill ecosystem. We then conduct a user study to validate our results and further answer these research questions. 

\paragraph{\textbf{Contributions.}} We make the following contributions:

    \textit{\underline{Automated System for Skill Analysis (RQ0).}} We present a natural-language-based
    system, ``\systemname,'' that automatically interacts with Alexa skills and collects their contents at scale (Section~\ref{section:system}).
    \systemname generates valid skill inputs, analyzes skill responses, and systematically generates follow-up inputs. 
    \systemname can be run longitudinally to identify new conversations and new conversation branches in previously analyzed skills. 
    We made our project available to the public to help facilitate future research.\footnote{\url{https://lenhattu.com/research/skillbot/}}
    
    \textit{\underline{Identification of Risks to Children (RQ1).}}
    We take a systematic approach to analyze VPA skills based on automated interactions.
    We analyze 3,434 Alexa skills specifically targeted toward children in order to measure the prevalence of kid skills that contain risky content, including inappropriate language and personal information collection (Section~\ref{section:skillanalysis}). 
    Through multiple rounds of interactions, we identify 28 kid skills with risky contents and maintain a growing dataset of 31,966 non-overlapping skill behaviors. 
    
    \textit{\underline{User Study of Parents' Awareness and Experiences (RQ2).}}
    We aim to better understand the real world contexts of children's interactions with VPAs and verify our \systemname results by seeing whether parents also viewed identified skills as risky. We conduct a user study of 232 U.S. Alexa users who have children under 13 years old (Section~\ref{section:userstudy}). A majority of surveyed parents express concern about the content of the risky kids skills identified by \systemname. Many also express disbelief that these skills are actually available for Alexa VPAs. This lack of risk awareness is compounded by findings that many parents do not use VPA parental controls and allow their children to use VPA versions that do not have parental controls enabled by default.
    
    \textit{\underline{Confounding Utterances (RQ3).}} We identify confounding utterances as a novel threat to VPA users (Section~\ref{section:utterances}). Confounding utterances are voice commands shared by more than one skill that may be present on a VPA device. When a user interacts with a VPA via a confounding utterance, it might trigger a reaction from any of these skills. If a kid skill shares a confounding utterance with a skill inappropriate for kids, a child user might inadvertently  interact with the inappropriate skill. For example, a child could use a confounding utterance to invoke a kid skill X only to have non-kid skill Y triggered instead. As many VPAs do not offer visual cues of what skill is actually invoked, the user may not realize that skill Y is running instead of skill X. 
    Our analysis reveals 4,487 confounding utterances shared between two or more skills and highlights those that place child users at risk of invoking a non-kid skill instead of a kid skill.
\section{Background}
\label{section:background}
In this section, we first provide an overview of what a voice personal assistant (VPA) is and how it works. We then describe how voice applications (i.e., skills) are implemented, published, and used.

\paragraph{\textbf{Voice Personal Assistant.}}
A VPA is a software agent that can interpret users' speech to perform certain tasks or answer questions from users via a synthesized voice. 
Most VPAs, such as Amazon Alexa and Google Home, follow a cloud-based system design. In particular, when the user speaks to the VPA device with a request, this request is sent to the VPA service provider's cloud server for processing and invoking the corresponding skills. Third-party skills can also be hosted on external web services instead of the VPA service provider's cloud server. 

\paragraph{\textbf{Building and Publishing Skills.}}
To provide a broader range of features, Amazon allows third parties to develop skills for Alexa via Alexa Skills Kit (ASK)~\cite{alexaskillskit}. Using ASK, developers can build custom Alexa skills that use their own web services to communicate with Alexa~\cite{customalexaskill}. There are currently more than 50,000 skills with a wide variety of features, including reading news, playing games, controlling a smart home, checking credit card balances, and telling jokes, that are publicly available on the Alexa Skills store~\cite{alexastore}.

\paragraph{\textbf{Enabling and Invoking Skills.}}
Unlike mobile apps, Alexa skills are hosted on Amazon's cloud servers. Therefore, users do not have to download binary files or run an installation process. To use a skill, users only need to enable it in their Amazon account. 
There are two ways to enable or disable a skill. The first is via the enable/disable button on a skill's info page. Users can access the skill info page via the Alexa Skills store on Amazon's website or the Alexa companion app. The second is via voice command. Note that for usability, Amazon also allows users to invoke skills directly through voice commands without needing to enable the skill first. 

Users can invoke a skill by saying one of its invocation phrases~\cite{alexainvokeskill}. Invocation phrases include two types: with intent and without intent. For example, a user could say ``Alexa, open Ted Talks'' to invoke the Ted Talks skill (without intent) or ``Alexa, open Daily Horoscopes for Capricorn'' to invoke the Daily Horoscopes skill and ask it to provide information about Capricorn (with intent). Since there are different ways of paraphrasing sentences, there are multiple variants of invocation phrases that perform the same task.

Alexa also allows some flexibility in invoking skills through the name-free interaction feature~\cite{alexanamefreeinteraction}. The user can speak to Alexa with a skill request that does not necessarily include the skill name. Alexa will process the request and select a top candidate skill that fulfills the request. If the chosen skill is not yet enabled by the user, it may be auto-enabled for the user. Every skill has an Amazon webpage, which includes at most three \textit{sample utterances}, i.e., voice commands with which users could verbally interact with the skill. In addition, the webpage may include an ``Additional Instructions'' section with additional example voice commands.

\section{Alexa Parental Control, Permission Control, and their Limitations}
In this section, we introduce the parental control and permission control features for protecting children using Alexa VPAs and discuss some of their limitations.

\paragraph{\textbf{Alexa Parental Control.}}
Amazon FreeTime is a parental control feature which allows parents to manage what content their children can access on their Amazon devices. FreeTime on Alexa provides a Parent Dashboard user interface for parents to set daily time limits, monitor activities, and manage allowed content. If Freetime is enabled, users can only use skills in the kids category by default. To use other skills, parents need to manually add those skills to a whitelist. FreeTime Unlimited is a subscription that offers child-friendly content, including a list of kid skills available on compatible Echo devices, for children under 13. Parents can purchase this subscription via their Amazon account and use it across all compatible Amazon devices.

Children can potentially access an Amazon Alexa device located in a shared space and invoke ``risky'' skills in the absence of child-protection features. FreeTime is turned off by default on the regular version of Amazon Echo, and previous studies, such as those in medicine~\cite{johnson2003defaults}, psychology~\cite{mckenzie2006recommendations}, and behavioral economics~\cite{madrian2001power}, have shown that people often opt for default settings. Although parents can turn on FreeTime,
the feature places a burden on adult users. For example, users sometimes cannot remove or disable skills added by FreeTime (which has been an issue since 2017~\cite{amazonforumfreetime, redditfreetime}). Some users find it hard to access the list of skills available via FreeTime Unlimited~\cite{redditfreetimeunlimited, amazonforumfreetimeunlimited}. Additionally, some skills that parents would like to use themselves may not be appropriate for kids and thus not allowed in FreeTime mode by default. This leads to users mistakenly thinking that not being able to use a skill in FreeTime mode is a bug of the skill itself, which leads to complaints being sent to the skill developer~\cite{plexcomplaint}. If parents want to use non-kid skills in FreeTime mode, they have to manually add these skills to the whitelist in the parent dashboard interface. Alternatively, they would have to remember to enable or disable FreeTime at appropriate times, negatively affecting the user experience.

\paragraph{\textbf{Alexa Permission Control.}}
Some Alexa skills need personal information from users to give accurate responses or to process transactions. Skills should formally request access to personal information such that when a user first enables the skill, Alexa asks the user to go to the Alexa companion app to grant the requested permission. 
However, this permission control mechanism only protects personal information in the user's Amazon Alexa account. If a skill does not specify permission requests, but directly asks for personal information through a voice interaction with the user, they can easily bypass this control.

\section{Threat Model}
\label{section:threatmodel}
In this paper, we consider two main types of threats: (1) risky skills (i.e., skills that contain inappropriate content or ask for users' personal information through voice interaction) and (2) confounding utterances (i.e., utterances that are shared among two or more different skills). This section explains these two threats. 

\paragraph{\textbf{Risky Skills.}}
We investigate skills containing risky content that may harm children. We define ``risky'' skills as containing either or both of two kinds of content: (1) inappropriate content for children or (2) requests for personal information through voice interaction. An example of inappropriate content is the ``My Burns'' skill in Amazon's Kids category that says ``You're so ugly you'd scare the crap out of the toilet.'' Another example is a kid skill named ``Shape Game'' that asks for age information: ``Awesome! Before we start however; I’m curious...how old are you?'' Note that these risks may be subjective. Thus, we evaluate our findings via a user study to understand the impacts of these risky skills. These threats may come from an adversary who intentionally develops malicious skills or a benign/inexperienced developer who is not aware of the risks.

\paragraph{\textbf{Confounding Utterances.}}
We identify a new risk which we call ``confounding utterances''. We define confounding utterances as voice commands that are shared among two or more different skills. Effectively, a confounding utterance could trigger an unexpected skill for the user.

Confounding utterances are different from previous research on voice squatting attacks, which exploited the speech recognition misinterpretations made by voice personal assistants~\cite{zhang2019lipfuzzer, kumar2019emerging, kumar2018squatting, zhang2019dangerousskills}. 
They showed that voice command misinterpretation problems due to spoken errors could yield unwanted skill interactions, and an adversary can route users to malicious Alexa skills by giving the malicious skills invocation names that are pronounced similarly to legitimate ones. 

In contrast, this paper considers a new risk that exists even if there is no such voice command misinterpretation: Alexa may invoke a skill that the user did not intend if multiple skills are configured to respond to identical (confounding) utterances. We want to find out, given a confounding utterance that is shared between multiple skills, which skill Alexa prioritizes to enable/invoke. This matters because users have no control over which skill is actually opened when Alexa recognizes an intentional or unintentional voice command containing the confounding utterance (e.g., if Alexa is triggered by background conversations). 
The risk of confounding utterances is further exacerbated by Alexa's name-free interaction feature~\cite{alexanamefreeinteraction}, which allows Alexa to invoke a skill even if a user does not explicitly state its invocation name. 
Furthermore, the lack of a download or installation process when a new skill is invoked makes it easy for these skills to bypass user awareness.
In summary, a confounding utterance may cause a user to invoke a skill other than the one they intended. For instance, a child may have one skill in mind but accidentally invoke a different skill that responds to confounding utterances. An adversary could therefore exploit confounding utterances to get children to use malicious skills.
\section{Automated Interaction with Skills}
\label{section:system}
To study the impacts that risky skills might have on children, we propose \systemname, which systematically interacts with skills to discover risky content and confounding utterances (\textbf{RQ0}). 
In this section, we first show how we design \systemname for interacting with skills and collecting their responses thoroughly and at scale. We then evaluate \systemname for its reliability, coverage, and performance. 

\begin{figure}[htbp]
  \centering
  \includegraphics[width=\linewidth]{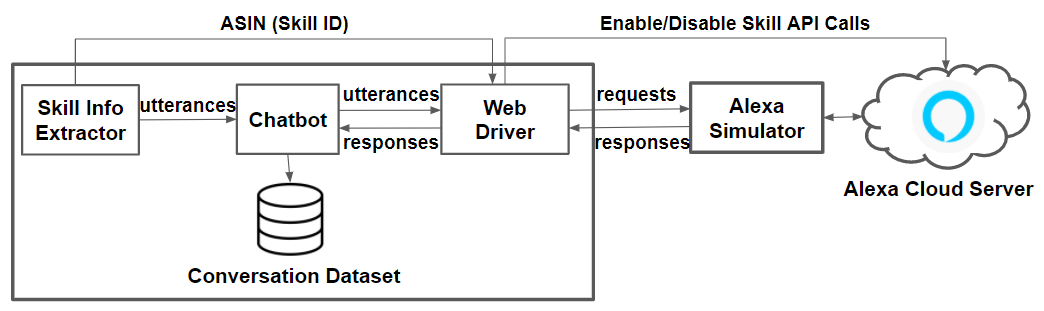}
  \caption{Automated skill interaction pipeline.}
  \label{fig:automationpipeline}
\end{figure}

\subsection{Automated Interaction System Design}
\label{system:design}
Our goal is for \systemname to interact effectively and efficiently with Alexa skills to uncover risky content for children in the skills' behaviors thoroughly and at scale.

\paragraph{\textbf{Overview.}}
Our system consists of four main components: (1) skill information extractor, (2) web driver, (3) chatbot, and (4) conversation dataset (Figure~\ref{fig:automationpipeline}). The skill information extractor handles exploring, downloading, and parsing information about skills available in the Alexa Skills Store. The web driver handles connections to Alexa and requests from/to the skills. The chatbot discovers interactions with the skills and records the conversations into the conversation dataset.

\paragraph{\textbf{Skill Information Extractor.}}
Amazon provides an online repository of skills via the Alexa Skills Store~\cite{alexastore}. Each skill is an individual product, which has its own product info page and an Amazon Standard Identification Number (ASIN) that can be used to search for the skill in Amazon's catalogue. The URL of a skill's info page can be constructed from its ASIN. Our skill information extractor includes a web scraper to systematically access the Alexa website and download the skills' info pages in HTML based on their ASINs. It then reads the HTML files and constructs a JSON dictionary structure using the BeautifulSoup library~\cite{beautifulsoup}. For each skill, we extract any information available on its info page, such as ASIN (skill ID), icon, sample utterances, invocation name, description, reviews, permission list, and category (e.g., kids, education, smart home, etc.).

\paragraph{\textbf{Web Driver.}}
We leverage Amazon's Alexa developer console~\cite{alexasimulator} to allow programmatic interactions with skills using text inputs. Our web driver module uses the Selenium framework, which is a popular web browser automation framework for testing web applications, to automate sending requests to Alexa and interacting with the skill info page to check the status of the skill (i.e., enabled, disabled, not available). We also implement a module that handles skill enabling/disabling requests. This module uses private APIs derived from inspecting XMLHttpRequests within network activities of Alexa webpages.

\paragraph{\textbf{Chatbot.}}
\label{system:chatbot}
Our NLP-based chatbot module interacts with the skills and explores as much of the skills' content as possible. The module includes several techniques to explore sample utterances suggested by the skill developers, create additional utterances based on the skill's info, classify utterances, detect questions in responses, and generate follow-up utterances. In the following paragraphs, we provide details of how our chatbot module explores and classifies utterances, detects questions, and generates follow-up utterances.

\begin{figure}[htbp]
  \centering
  \includegraphics[width=\linewidth]{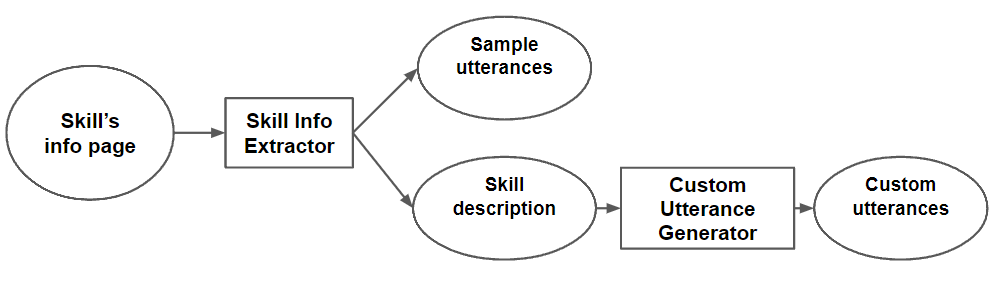}
  \caption{Workflow of exploring potential utterances to test a skill. Our chatbot module collects a variety of utterances, including sample utterances from the skill's info page and custom utterances constructed based on the skill description.}
  \label{fig:exploreutterances}
\end{figure}

\textit{\underline{Exploring and Classifying Utterances:}} Amazon allows developers to list up to three sample utterances on their skill's information page. Our system first extracts these sample utterances. Some developers also put additional instructions into their skill's description. Therefore, our system further processes the skill's description to generate more utterances. In particular, we consider sentences that start with an invocation word (i.e., ``Alexa, ...'') to be utterances. We also notice that phrases inside quotes can be utterances. An example is ``You can say `give me a fun fact' to ask the skill for a fun fact.'' Once a list of collected utterances is constructed, our system classifies these utterances into opening and in-skill utterances. Opening utterances are used to invoke/open a skill. These often include the skill's name and start with opening words such as ``open,'' ``launch,'' and ``start''~\cite{alexainvokeskill}. In-skill utterances are used within the skill's session (when the skill is already invoked). Some examples include ``tell me a joke,'' ``help,'' or ``more info.'' Figure~\ref{fig:exploreutterances} shows the workflow of how our chatbot module prepares potential utterances to test a skill. 

\begin{figure}[htbp]
  \centering
  \includegraphics[width=\linewidth]{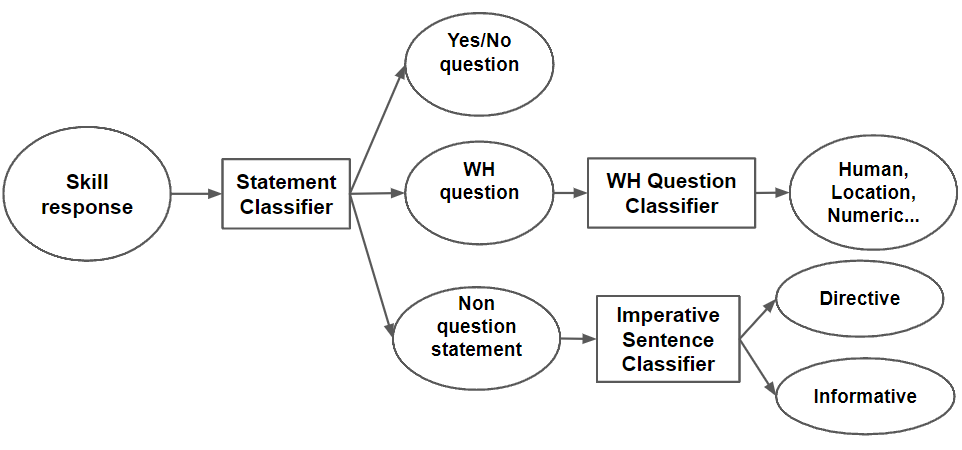}
  \caption{Our chatbot module's workflow for classifying a skill's responses in order to generate follow-up utterances.}
  \label{fig:skillresponseclassifier}
\end{figure}

\textit{\underline{Detecting Questions in Skill Responses:}} To extend the conversation, our system first classifies responses collected from the skill into three main categories: yes/no questions, WH questions, and non-question statements. For this classification task, we employ spaCy~\cite{spacy} and StanfordCoreNLP~\cite{manning2014stanfordnlp, qi2018universal}, which are popular tools for NLP tasks. In particular, we first tokenize the skill's response into sentences and each sentence into words. We then annotate each sentence using part-of-speech (POS) tagging, including both TreeBank POS tags~\cite{taylor2003treebanks} and Universal POS tags~\cite{universalpostags}. The POS tags allow us to identify the role of each word in the sentence (e.g., auxiliary, subject, or object). A yes/no question usually starts with an auxiliary verb, which follows the subject-auxiliary inversion formation rule. Yes/no questions generally take the form of [auxiliary + subject + (main verb) + (object/adjective/adverb)?]. Some examples are ``Is she nice?'' ``Do you play video games?'' and ``Do you swim today?'' It is also possible to have the auxiliary verb as a negative contraction such as ``Don't you know it?'' or ``Isn't she nice?'' A WH question contains WH words such as ``what,'' ``why,'' or ``how.'' These WH words can be identified in the sentence based on their POS tags: WDT, WP, WP\$, and WRB. Regular WH questions usually take the form of [WH-word + auxiliary + subject + (main verb) + (object)?]. Some examples are ``What is your name?'' and ``What did you say?'' Furthermore, we consider pied-piping WH questions such as ``To whom did you send it?'' We exclude cases in which WH words are used in a non-question statement, such as ``What you think is great,'' ``That is what I did,'' or ``What goes around comes around.'' Figure~\ref{fig:skillresponseclassifier} shows our chatbot module's workflow for categorizing responses from skills. The chatbot then continues the conversations with the skills by generating follow-up utterances based on these categories.

\textit{\underline{Generating Follow-up Utterances:}} Given a skill response, our chatbot follows up based on the category of the response.
(1) Yes/no questions. These questions ask for confirmation from the user, expecting either a ``yes'' or a ``no'' answer. The chatbot sends ``yes'' or ``no'' as a follow-up utterance to continue the conversation. (2) WH questions. The chatbot responds to WH questions based on the theme of the question. We use the classification method presented in~\cite{madabushi2016question} to categorize WH questions into one of six themes: abbreviation, entity, description, human, location, and numeric~\cite{questionclasses}. Abbreviation questions ask about a short form of an expression (e.g., ``What is the abbreviation for California?''). Entity questions ask about objects that are not human (e.g., ``What is color is the sky?''). Description questions ask about explanations of concepts (e.g., ``What does a defibrillator do?''). Human questions ask about individuals or groups of people. Location questions ask about places, such as cities, countries, states, etc. Numeric questions ask about numerical values, such as count, weight, size, etc. Each question theme can also be divided into subthemes. Our chatbot contains a dictionary of answers to specific subthemes (e.g., ``human:age'': [1, 2, 3, ...], ``location:states'': [Oregon, Arizona, ...]) that can be used to continue the conversation with the skill. For questions in subthemes that are too general (including many in the description theme), the chatbot replies with ``I don't know. Please tell me.'' to prompt for further responses from the skill.
(3) Non-question statements. These include two types of statements: directive statements and informative statements. Some directive statements ask the user to provide an answer, which is similar to a WH question (e.g., ``Please tell us your birthday''). For these cases, the chatbot parses the sentence to identify what being asked and responds as it would to a WH question. Other directive statements suggest words/phrases for the user to select to continue the conversation. Some examples include ``Please say `continue' to get a fun fact'' and ``Say `1' to get info about a book or `2' to get info about a movie.'' In these cases, the chatbot extracts the suggested words/phrases and uses them to continue the conversation. Informative statements provide users with some information, such as a joke, a fact, or daily news. These statements often do not indicate what the user can say to continue the conversation. The chatbot therefore sends an in-skill utterance ``Tell me another one'' or ``Tell me more'' as a follow-up to explore more content from the skill.

\paragraph{\textbf{Conversation Dataset.}}
\label{system:dataset}
Our conversation dataset is a set of JSON files, each of which represents a skill. The files contain a list of conversations with each skill collected by the chatbot module. Each conversation is stored as a list in which even indexes are the utterances sent by \systemname and odd indexes are the corresponding responses from the skill.

\subsection{Skill Conversation Data Collection}
We ran SkillBot to collect conversation data from Alexa skills at scale. 
We explain \systemname's detailed workflow for this process in the following paragraphs.

For each skill, \systemname conducts multiple round of interactions to explore different \textit{paths} within the \textit{conversation tree}. Each node in this tree is a unique response from Alexa. There is an edge between nodes $i$ and $j$ if there exists an interaction in which Alexa says $i$, the user (i.e., \systemname) says something, and then Alexa says $j$. We call the progression from $i$ to $j$ a \textit{path} in the tree. Multiple paths of interactions can exist for a skill. For instance, node $i$ could have two edges: one with $j$ and another one with $k$. Effectively, two paths lead from $i$. In one path, the user says something after hearing $i$, and Alexa responds with $j$. In another path, the user says something else after hearing $i$, and Alexa responds with $k$.

\begin{figure}[htbp]
  \centering
  \includegraphics[width=.7\columnwidth]{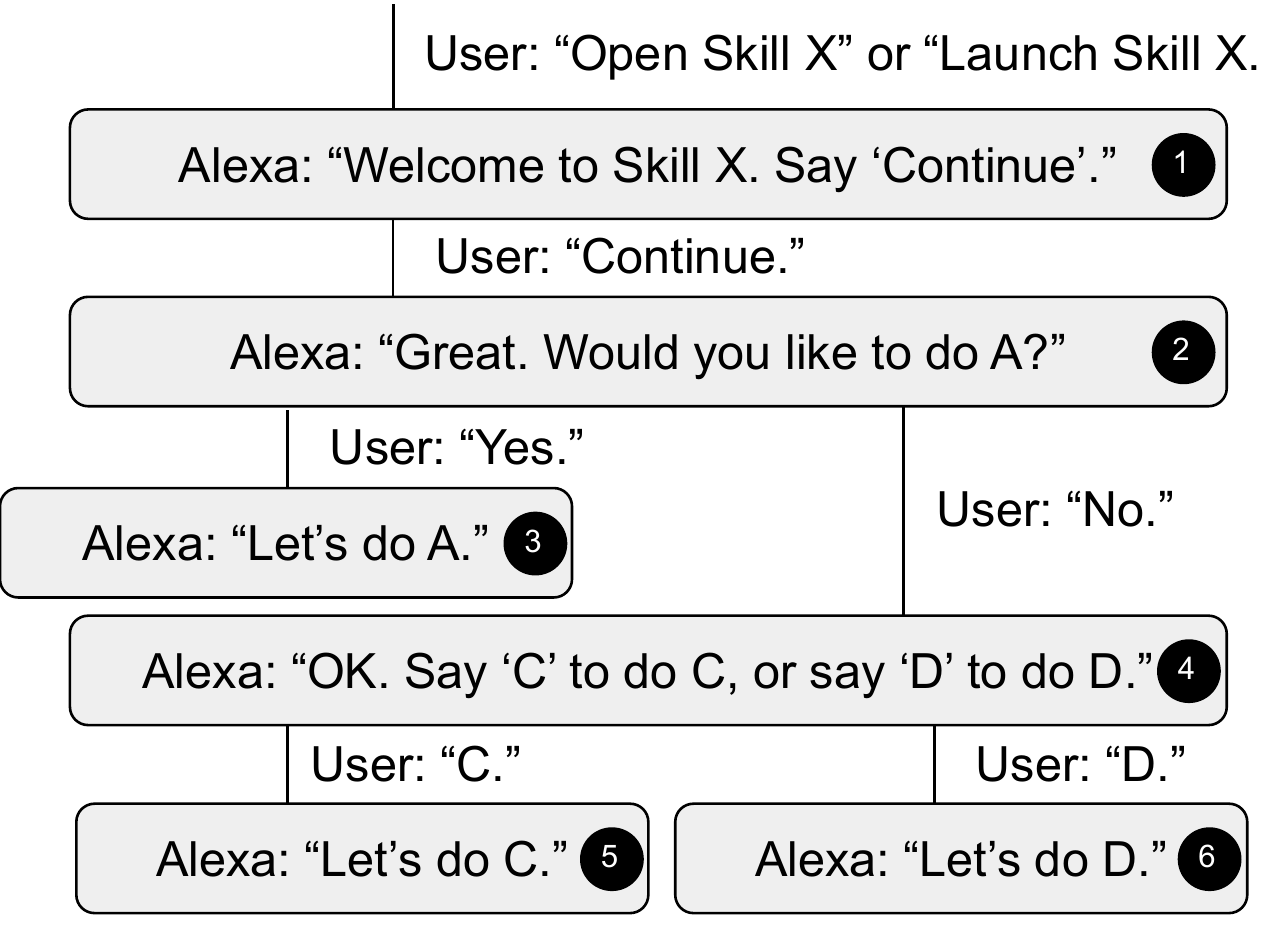}
  \caption{A conversation tree that represents how \systemname interacts with a typical skill.}
  \label{fig:branches}
\end{figure}

To illustrate how we construct a conversation tree for a typical skill, we show a hypothetical example in Figure~\ref{fig:branches}. First, the user would launch a skill by saying ``Open Skill X'' or ``Launch Skill X.'' This initial utterance could be found in the ``Sample Utterances'' section of the skill's information page; alternatively, it could also be displayed in the ``Additional Instructions'' section on the skill's page. Following the example in Figure~\ref{fig:branches}, let us assume that either ``Open Skill X'' or ``Launch Skill X'' triggers the same response from Alexa: ``Welcome to Skill X. Say `Continue'.'' This response is denoted by node~1 in Figure~\ref{fig:branches}. The user would say ``Continue'' and trigger another response from Alexa (node~2): ``Great. Would you like to do A?'' The user could either respond with ``Yes,'' which would trigger the response in node~3, or ``No,'' which would trigger node~4.

SkillBot explores multiple paths of the conversation tree by interacting with a skill multiple times, picking a different response each time. Following the example in Figure~\ref{fig:branches}, SkillBot could follow the path along nodes 1, 2, and 3 the first time it evaluates this skill. Once at node 3, the skill does not provide the user with the option to return to node 2, so SkillBot would have to start over to explore a different path. In the second run, SkillBot could follow a path along nodes 1, 2, 4, and 5. SkillBot responds with ``No'' after node 2 because it remembers answering ``Yes'' in the previous run. In the third run, SkillBot could follow nodes 1, 2, 4, and 6.

Each run of SkillBot terminates when exploring down a particular path is unlikely to trigger new responses from Alexa; in this case, SkillBot starts over with the same skill and explores a different path. We list four conditions where SkillBot would terminate a particular run: (1) Alexa's response is not new; in other words, SkillBot has seen the response repeatedly. SkillBot's goal is to maximize the interaction with unique skill responses in order to discover risky contents. (2) Alexa's response is empty. (3) Alexa's response is a dynamic audio clip (e.g., music or podcast) that does not rely on Alexa's automated voice. Due to limitations of the Alexa simulator, SkillBot is unable to extract and parse dynamic audio clips. Therefore, SkillBot terminates a path if it sees a dynamic audio clip because it does not know how to react. (4) Alexa's response is an error message, such as ``The service is unavailable.'' or ``Sorry, I don't understand.''

\subsection{Evaluation}
\label{system:evaluation}
In this section, we present our validation to ensure that interacting with skills via \systemname (Section~\ref{section:system}) can represent a user's interaction with skills via a physical Echo device. 

\paragraph{\textbf{Interaction Reliability}}
\label{system:evaluation:simulator}
We randomly selected 100 skills for validation. We used an Echo Dot device to manually interact with the skills and compared the responses against those collected by Skillbot. 
If the responses did not match, we further checked the skill invocation in the Alexa activity log to see if the same skill was invoked. We found that Skillbot and the Echo Dot have similar interactions across 99 of the 100 selected skills. Among these 99 skills, two skills responded with audio playbacks, which are not supported by the Alexa developer console~\cite{alexasimulatorlimit} employed by \systemname (Section~\ref{section:discussion}). However, their invocations were shown in the activity log, which matched those invocations when using the Echo Dot. We cannot verify the one remaining skill as Alexa cannot recognize its sample utterances, potentially due to an issue with the skill's web service.

\paragraph{\textbf{Skill Response Classification.}}
As described in Section~\ref{system:chatbot}, \systemname extends the conversation with a skill by classifying the skill's responses as yes/no questions, WH questions, and non-question statements.  To evaluate the performance of this classification, we randomly sampled 300 unique skill responses from our conversation collection and manually labeled them as ground truth. This ground truth set included 52 yes/no questions, 50 WH questions, and 198 non-question statements. We then used \systemname to label these responses and verified the labels against the ground truth. \systemname predicted 56 yes/no questions, 50 WH questions, and 194 non-question statements, which is over 95\% accuracy. The performance details for each class is shown in Table~\ref{tab:classificationperformance} (see Table~\ref{tab:classificationmatrix} in Appendix~\ref{appendix:classificationmatrix} for the confusion matrix of our 3-class classifier).

\begin{table}[htbp]
\centering
\begin{tabular}{|l|c|c|c|c|}
\hline
 & Accuracy & Precision & Recall & F1 Score \\ \hline
Yes/No & 98\% & 0.91 & 0.98 & 0.94 \\ \hline
WH question & 98\% & 0.94 & 0.94 & 0.94 \\\hline
Non-question & 96\% & 0.98 & 0.96 & 0.97 \\ \hline
\end{tabular}
\caption{Skill response classification performance}
\label{tab:classificationperformance}
\end{table}

\begin{figure*}[htbp]
  \centering
  \includegraphics[width=\textwidth]{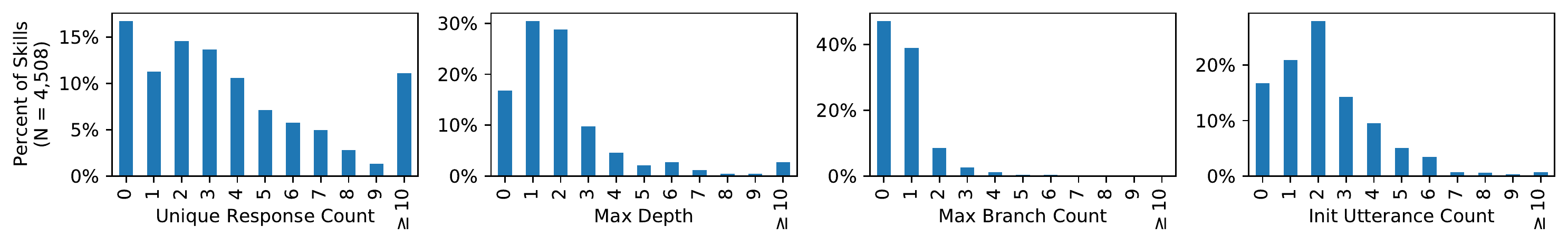}
  \caption{Coverage of SkillBot in terms of four criteria: number of unique responses from Alexa; maximum depth in a conversation tree; maximum number of branches for any node in a conversation tree; and number of initial utterances.}
  \label{fig:coverage}
\end{figure*}

\paragraph{\textbf{Coverage.}} We measure SkillBot's coverage by analyzing the collected conversation trees for every skill. Each skill can have multiple conversation trees representing different conversations. Our analysis includes four criteria: (1) the number of unique responses from Alexa, i.e., the number of nodes in a tree; (2) the maximum depth (or height) in a tree; (3) the maximum number of branches in a tree, i.e., how many options that SkillBot explored; and (4) the number of initial utterances, which counts the number of distinct ways to start interacting with Alexa. We show the results in Figure~\ref{fig:coverage}.

Per the second chart in Figure~\ref{fig:coverage}, SkillBot is able to reach a conversation tree depth of at least 10 on 2.7\% of the skills. Such a depth allows SkillBot to trigger and explore a wide variety of Alexa's responses from which to discover risky contents. In fact, out of the 28 risky kid skills we identify (Section~\ref{section:skillanalysis}), 2 skills were identified at depth 11, 1 skill at depth 5, 4 skills at depth 4, 6 at depth 3, 8 at depth 2, and 7 at depth 1.

Per the fourth chart in Figure~\ref{fig:coverage}, SkillBot is able to initiate conversations with skills using more than 3 different utterances. Normally, a skill's information page lists \textit{at most} three sample utterances. In addition to using these sample utterances, SkillBot also discovers and extracts utterances in the ``Additional Instructions'' section on the skill's page. As a result, SkillBot interacted with 20.3\% of skills using more than 3 utterances. These extra initial utterances allow SkillBot to trigger more responses from Alexa. As we will explain in Section~\ref{section:skillanalysis}, 3 out of the 28 risky kid skills were discovered by SkillBot from additional utterances not listed among the sample interactions. 

\paragraph{\textbf{Time Performance.}}
It took about 21 seconds on average to collect one conversation. \systemname interacted with 4,507 skills and collected 39,322 conversations within 46 hours using five parallel processes on an Ubuntu 20.04 machine with an Intel Core i7-9700K CPU.

\section{Kid Skill Analysis}
\label{section:skillanalysis}
To investigate the risks of skills specifically targeted to children (\textbf{RQ1}), we employed \systemname to collect and analyze 31,966 conversations from a sample of 3,434 Alexa kid skills. In this section, we describe our dataset of kid skills and present our findings of risky kid skills. 

\subsection{Dataset}
\label{analysis:dataset}
Our system first downloaded information about skills from their info pages on the Alexa U.S.~skills store. We filtered out error pages (e.g., 404 not found) after three retries as well as non-English skills. As a result, we collected 43,740 Alexa skills from 23 different skill categories (e.g., business \& finance, social, kids, etc.). Our system then parsed data about the skills, such as ASIN (i.e., skill's ID), icon, sample utterances, invocation name, description, reviews, permission list, and category, from the downloaded skill info pages.

For our analysis, we investigated all skills in Amazon's Kids category (3,439 kid skills). We ran \systemname to interact with each skill and record the conversations. To speed up the task, we ran five processes of \systemname simultaneously. Note that \systemname can be run multiple times for each skill to cumulatively collect new conversations and new conversation branches for that skill. As a result, our sample had 31,966 conversations from 3,434 kid skills after removing five skills that resulted in errors or crashed Alexa.

\subsection{Risky Kid Skill Findings}
\label{analysis:riskykidskills}
We performed content analysis on the conversations collected from 3,434 kid skills to identify risky kid skills that have inappropriate content or ask for personal information. \revise{Examples of risky skills identified in our analysis are attached in Appendix~\ref{skillexamples}.}

\paragraph{\textbf{Skills with Inappropriate Content for Children.}}
Our goal was to analyze the skills' contents to identify risky skills that provide inappropriate content to children. To identify inappropriate content, we combined WebPurify and Microsoft Azure's Content Moderator, which are two popular content moderation services providing inappropriate content filtering for websites and applications with a focus on child protection~\cite{webpurifychildren, azurecontentmoderator}. We implemented a content moderation module for \systemname in Python 3, leveraging the WebPurify API and the Azure Moderation API to flag skills that have inappropriate content for children.

The Skillbot content moderation module flagged 33 potentially risky skills with expletives in their contents. However, a human review process was necessary to verify this output, because whether or not a flagged skill's content is actually inappropriate for children depends on context. For example, some of the expletives (such as ``facial'' and ``sex'') are likely appropriate in some conversational contexts. For the human review process, four researchers on our team---who come from 3 countries (including the USA), all of whom are English speakers, and whose ages range from 22 to 35---independently reviewed each of the flagged skills and voted whether the skills' content is inappropriate for children. Skills that received three or four votes were considered to actually be inappropriate. Using this approach, we identified 8 kid skills with inappropriate content. \revise{Some examples include the ``New Facts'' skill that said ``Here's your fact: A pig's orgasm lasts for 30 minutes'' and the ``My Burns'' skill that said ``You're so ugly you'd scare the crap out of the toilet.''}
Out of these 8 kid skills, \systemname identified the inappropriate content of one skill at conversation tree depth 11, one at depth 5, two at depth 4, one at depth 2, and three at depth 1.

We also sampled 100 other skills that were not flagged as having inappropriate content by \systemname and manually checked them. We did not find inappropriate content in any of these skills, further validating Skillbot's performance. 

\paragraph{\textbf{Skills Collecting Personal Information.}}
Our goal was to detect if kid skills asked users for personal information. To the best of our knowledge, available tools only focus on detecting personal information in the given text input, which is a different goal. For this analysis, we employed a keyword search to identify skill responses that asked for personal information. We constructed a list of personal information keywords based on the U.S. Department of Defense Privacy Office~\cite{dodprivacy} and searched for these keywords in the skill responses. The keywords include name, age, address, phone number, social security number, passport number, driver's license number, taxpayer ID number, patient ID number, financial account number, credit card number, date of birth, and zipcode. However, a naive keyword search would not be sufficient, because in some cases the sentence containing any of those keywords might not actually ask for such information. We therefore combined keyword search and our SkillBot's question detection technique (presented in Section~\ref{system:design}) to detect if a skill asked the user to provide personal information.

22 risky kid skills were flagged as asking users for personal information. To verify the result, we manually checked these 22 skills and 100 random skills that were not flagged. The manual verification found 2 false positives and 0 false negatives. Thus, 20 kid skills asked for personal information, such as name, age, and birthday. Some examples include the ``Ready Freddy'' skill that kept asking kids to introduce themselves or the ``Birthday Wisher'' skill that asked for birthday info. Table~\ref{tab:risky-personal-info-confusion-matrix} in Appendix~\ref{appendix:riskyidentificationmatrix} presents the confusion matrix for the evaluation.

Out of these 20 skills, SkillBot identified the queries for sensitive information in one skill at conversation tree depth 11, two skills at depth 4, six at depth 3, seven at depth 2, and four at depth 1. SkillBot  also identified these queries in non-sample utterances for three of the skills (i.e., utterances listed in the ``Additional Instructions''  section of the skill's info page on Amazon.com rather than as sample interactions with the skill).

We further analyzed the permission requests of these 20 risky kid skills and found that none requested any permissions from the user.

\section{Awareness \& Opinions of Risky Kid Skills}
\label{section:userstudy}
To evaluate how the risky kid skills we identified actually impact child users (\textbf{RQ2}), we conducted a user study of 232 U.S. parents who use Amazon Alexa and have children under 13. Our goal was to qualitatively understand parents' expectations and attitudes about these risky skills, parents' awareness of parental control features, and how risky skills might affect children. 
Our study protocol was approved by our Institutional Review Board (IRB). The full text of our survey instrument is provided in Appendix~\ref{appendix:questionaire}. In this section, we describe our recruitment strategy, survey design, response filtering, and results.

\subsection{Recruitment}
\label{study:recruitment}
We recruited participants on Prolific,\footnote{https://www.prolific.co/} a crowdsourcing website for online research. Participants were required to be adults 18 years or older who are fluent in English, live in the U.S. with their children under 13, and have at least one Amazon Echo device in their home.
We combined Prolific's pre-screening filters and a screening survey to get this niche sample of participants for our main survey. Our screening survey consisted of two questions to determine (1) if the participant has children aged 1--13 and (2) if the participant has Amazon Echo device(s) in their household. 1,500 participants took our screening survey and 258 qualified for our main survey. The screening survey took less than 1 minute to complete and the main survey took an average of 6.5 minutes (5.2 minutes in the median case). Participants were compensated \$0.10 for completing the screening survey and \$2 for completing the main survey. \revise{To improve response quality, we limited both the screening and main surveys to Prolific participants with at least a 99\% approval rate.} 

\subsection{Screening Survey}
The screening survey consisted of two multiple-choice questions: ``Who lives in your household?'' and ``Which electronic devices do you have in your household?'' This allowed us to identify participants with children aged 1--13 and Amazon Echo device(s) in their household who were eligible to take the main survey.

\subsection{Main Survey}
The main survey consisted of the following four sections.

\paragraph{\textbf{Parents' Perceptions of VPA Skills.}}
This section investigated parents' opinions of and experiences with risky skills. Participants were presented with two conversation samples collected by \systemname from each of the following categories (six samples total). Conversation samples were randomly selected from each category for each participant and were presented in random order. 

\begin{itemize}
    \item \textit{Expletive.} Conversation samples from 8 skills identified in our analysis that contain inappropriate language content for children. 
    \item \textit{Sensitive.} Conversation samples from 20 skills identified in our analysis that ask the user to provide personal information, such as name, age, and birthday. 
    \item \textit{Non-Risky.} Conversation samples from 100 skills that did not contain inappropriate content for children or ask for personal information. 
\end{itemize}

The full list of skills in the Expletive and Sensitive categories are provided in Appendix~\ref{skillexamples}.
Each participant was asked the following set of questions after viewing each conversation sample:
\begin{itemize}
    \item Do you think the conversation is possible on Alexa?
    \item Do you think Alexa should allow this type of conversation?
    \item Do you think this particular skill or conversation is designed for families and kids?
    \item How comfortable are you if this conversation is between your children and Alexa?
    \item If you answered ``Somewhat uncomfortable'' or ``Extremely uncomfortable'' to the previous question, what skills or conversations have you experienced with your Alexa that made you similarly uncomfortable?
\end{itemize}

\paragraph{\textbf{Amazon Echo Usage.}} We asked which device model(s) of Amazon Echo our participants have in their household (e.g., Echo Dot, Echo Dot Kids Edition, Echo Show). We also asked whether their kids used Amazon Echo at home.

\paragraph{\textbf{Awareness of Parental Control Feature.}} We asked the participants if they think Amazon Echo supports parental control (yes/no/don't know). Participants who answered ``yes'' were further asked to identify the feature's name (free-text response) and if they used the feature (yes/no/don't know).

\paragraph{\textbf{Demographic Information.}} At the end of the survey, we asked demographic questions about gender, age, and comfort level with computing technology. Our sample consisted of 128 male (55.2\%), 103 female (44.4\%), and 1 preferred not to answer (0.4\%). The majority (79.7\%) were between 25 and 44 years old. Most participants in our sample are technically savvy (68.5\%). Table~\ref{tab:demographic} shows detailed demographic information of our participants.

\begin{table}[!htbp]
\centering
\small
\begin{tabular}{|lll|}
\hline
     & \textbf{Responses} & \textbf{Percentage}  \\ \hline
\multicolumn{3}{|l|}{\textbf{Gender}}                                    \\
Male                                         & 128 & 55\%        \\ 
Female                                       & 103   & 44\%           \\ 
Prefer not to answer                         & 1 & $<$1\%          \\ \hline
\multicolumn{3}{|l|}{\textbf{Age}}                                       \\ 
18 - 24                                      & 3     &  1\%        \\ 
25 - 34                                      & 61   &     26\%       \\ 
35 - 44                                      & 124  &    53\%         \\ 
45 - 54                                      & 40     & 17\%         \\ 
55 - 64                                      & 3    &  1\%          \\ 
65 and above                                 & 1    &   $<$1\%       \\ \hline
\multicolumn{3}{|l|}{\textbf{Comfort level with computing technology}}   \\ 
Ultra Nerd                                   & 19    &     9\%      \\ 
Technically Savvy                            & 159   &   68\%        \\ 
Average User                                 & 54    &    23\%       \\ 
Luddite                                      & 0    &     0\%       \\ 
\hline
\end{tabular}%
\caption{Demographic information (gender, age, comfort level with computing technology) of the participants in our sample. The ``Responses'' column contains the number of participants who selected the corresponding choices. Our sample is nearly gender-balanced with most participants in the 25--44 age group. Most participants also self-reported to be technically savvy or average users.}
\label{tab:demographic}
\end{table}

\subsection{Response Filtering}
\label{userstudy:response-filtering}
We received 237 responses for our main survey. We filtered out responses from participants who incorrectly answered either of two attention check questions (``What is the company that makes Alexa?'' and ``How many buttons are there on an Amazon Echo?''). We further reviewed the submissions and excluded participants who gave meaningless responses (e.g., straight-lining or entering only whitespaces into all free-text answer boxes). This resulted in 232 valid responses for analysis.

\subsection{User Study Results} \label{sec:survey-results}
We find that most parents allow their children to use types of Amazon Echo devices other than the Kids Edition. Such types of Echo devices do not have parental control enabled by default. We also find that many parents do not know about the parental control feature. For those who know about the feature, only a few of them use it. Thus, many children potentially have access to risky skills. Our results further show that parents are not aware of the risky skills that are available in the Kids category on Amazon. When presented with examples of risky kid skills that have expletives and those that ask for personal information, parents express concerns, especially for skills with expletives. Some parents reported previous experiences with such risky skills.

\paragraph{\textbf{Parents' Perceptions of Kid Skills.}}
Table~\ref{tab:yes_no_distribution} shows the distribution of responses to the following questions across the Expletive, Sensitive, and Non-Risky skill sets:
\begin{itemize}
   \item Do you think the conversation is possible on Alexa?
   \item Do you think Alexa should allow this type of conversation?
   \item Do you think this particular skill or conversation is designed for families and kids?
\end{itemize}
A majority of parents thought that the interactions with the Expletive skills were not possible and should not be allowed by Alexa. Only 45.9\% of the parents thought these interactions were possible and only 41.6\% thought such skills should be allowed. Furthermore, the majority of parents did not think that the Expletive skills were designed for families and kids.

The parents' responses with regard to the Expletive skills are significantly different from their responses to the Sensitive and Non-Risky skills on these questions. For each of these three questions, we conduct Chi-square tests on the pairs of responses across the skill sets: Non-Risky vs. Expletive, Non-Risky vs. Sensitive, and Expletive vs. Sensitive. The responses from the Expletive set are significantly different from responses from the other two sets for all three questions ($p < 0.05$). 
The responses to the ``Alexa should allow'' question are also significantly different for the Non-Risky set versus for the Sensitive set  ($p < 0.05$). 
In contrast, the responses for the ``Possible on Alexa'' and ``Designed for families and kids'' questions display no significant difference between Sensitive and Non-Risky sets. This is alarming, as the Sensitive skills ask for personal information through the conversations with users, thereby bypassing Amazon's built-in permission control model for skills. As many skills are hosted by third parties, sensitive information about children could be leaked to someone other than Amazon. 

\begin{table*}[htbp]
\centering
\small
\begin{tabularx}{\textwidth}{X | r r r| r r r | r r r}
  \toprule
    ~ & 
    \multicolumn{3}{c|}{Possible on Alexa} & 
    \multicolumn{3}{c|}{Alexa should allow} & 
    \multicolumn{3}{c}{Designed for families and kids} \\
    Response & 
    Non-Risky & Expletive & Sensitive &
    Non-Risky & Expletive & Sensitive &
    Non-Risky & Expletive & Sensitive  \\
    \midrule
    Yes & 78.0\% & 45.9\% & 71.6\% & 83.2\% & 41.6\% & 66.2\% & 68.1\% & 27.4\% & 55.8\% \\ 
No & 7.5\% & 30.2\% & 11.2\% & 8.8\% & 44.2\% & 16.6\% & 13.8\% & 57.1\% & 16.6\% \\ 
Not sure & 14.4\% & 23.9\% & 17.2\% & 8.0\% & 14.2\% & 17.2\% & 18.1\% & 15.5\% & 27.6\% \\ 

  \bottomrule
\end{tabularx}
\caption{Distribution of responses for each of the three yes/no questions across three types of conversations.}
\label{tab:yes_no_distribution}
\end{table*}

\paragraph{\textbf{Designed for Family and Children.}}
Table~\ref{tab:family_oriented} shows the distribution of responses for the question: ``Do you think this particular skill or conversation is designed for families and kids?'' with a breakdown across different types of skills (e.g., Non-Risky, Expletive, and Sensitive). 
Most parents (57.1\%) felt that skills with expletives were not designed for families and kids. In addition, 15.5\% were not sure if such skills were designed for families and kids. These results indicate that the parents were not aware of the skills with expletives that were \textit{actually} developed for kids and published in Amazon's ``Kids'' category. In addition, about half of parents (44.2\%) did not think the Sensitive skills were designed for families/kids, although these skills are actually in the ``Kids'' category on Amazon as well.

\begin{table}[t]
\centering
\small
\begin{tabularx}{0.4\textwidth}{X | r| r| r r }
  \toprule
    Response & Expletive & Sensitive & Non-Risky \\
    \midrule
    Yes & 27.4\% & 55.8\% & 68.1\% \\ 
No & 57.1\% & 16.6\% & 13.8\% \\ 
Not sure & 15.5\% & 27.6\% & 18.1\% \\ 

  \bottomrule
\end{tabularx}
\caption{Distribution of responses for the question: ``Do you think this particular skill or conversation is designed for families and kids?'' }
\label{tab:family_oriented}
\end{table}

\begin{figure}[htbp]
  \centering
  \includegraphics[width=\columnwidth]{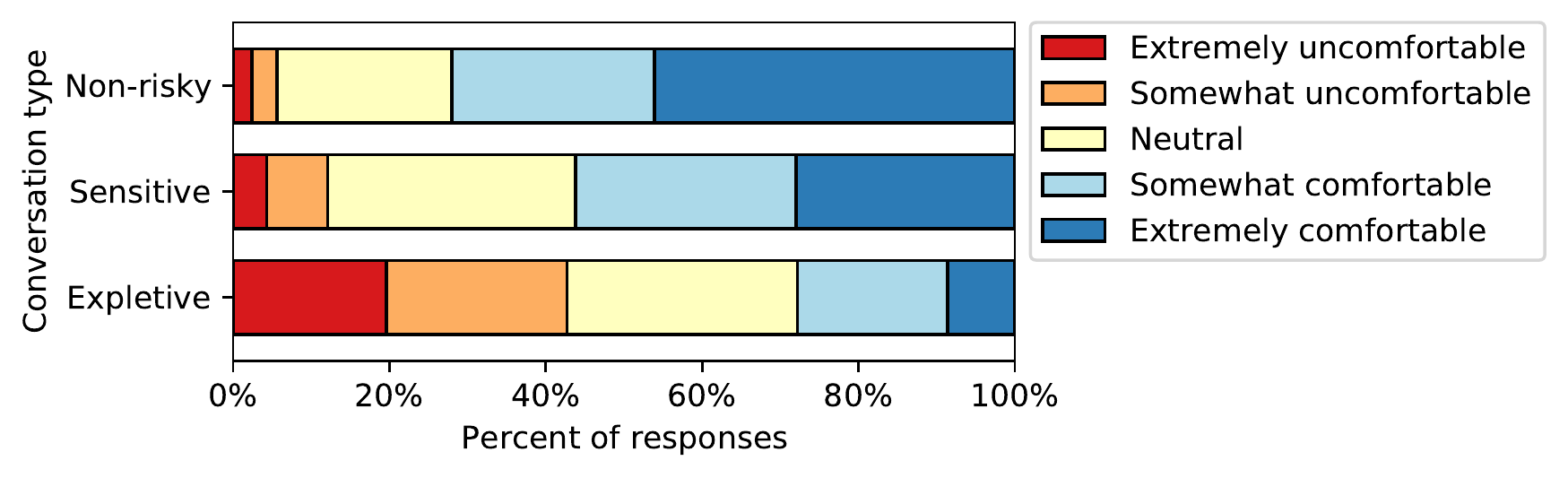}
  \caption{Participants' comfort levels if conversations of a particular type were to happen between the participants' children and Alexa.}
  \label{fig:comfortable_with_kids}
\end{figure}

\paragraph{\textbf{Parents' Comfort Level.}}
We used a five-point Likert scale to measure parents' comfort levels if the presented conversations were between their children and Alexa. Figure~\ref{fig:comfortable_with_kids} shows the participants' comfort levels for each skill category. These results indicate that parents were more uncomfortable with the Expletive skill conversations compared to the Sensitive skill conversations. In particular, 42.7\% of the respondents expressed discomfort (``Extremely uncomfortable'' and ``Somewhat uncomfortable'') with the Expletive skills, compared to only 12.1\% with the Sensitive skills and 5.6\% with the Non-Risky skills.

Chi-square tests show that parents' comfort with the Expletive conversations is significantly different from their comfort with the Sensitive conversations and from the Non-Risky conversations ($p < 0.05$).
Some participants expressed their concerns about skills in the Expletive set by free-text responses, including ``It doesn't seem appropriate to tell jokes like this to children (P148),'' ``Under no circumstances should anyone have a coversation [sic] with children about orgasms.  This would be grounds for legal action (P163),'' ``I do not believe Alexa should be used in such a crass manner or to teach my child how to be crass (P210),'' ``Poop and poopy jokes don't happen in my household (P216),'' and ``It is too sexual (P123).'' Beyond the skills shown in the survey, one respondent also recalled hearing similar skills such as ``Roastmaster (P121).'' Another respondent remembered something similar but was unable to provide the name of the skill: ``We have asked Alexa to tell us a joke in front of our young son and Alexa has told a few jokes that were borderline inappropriate (P140).'' 

\begin{figure}[htbp]
  \centering
  \includegraphics[width=\columnwidth]{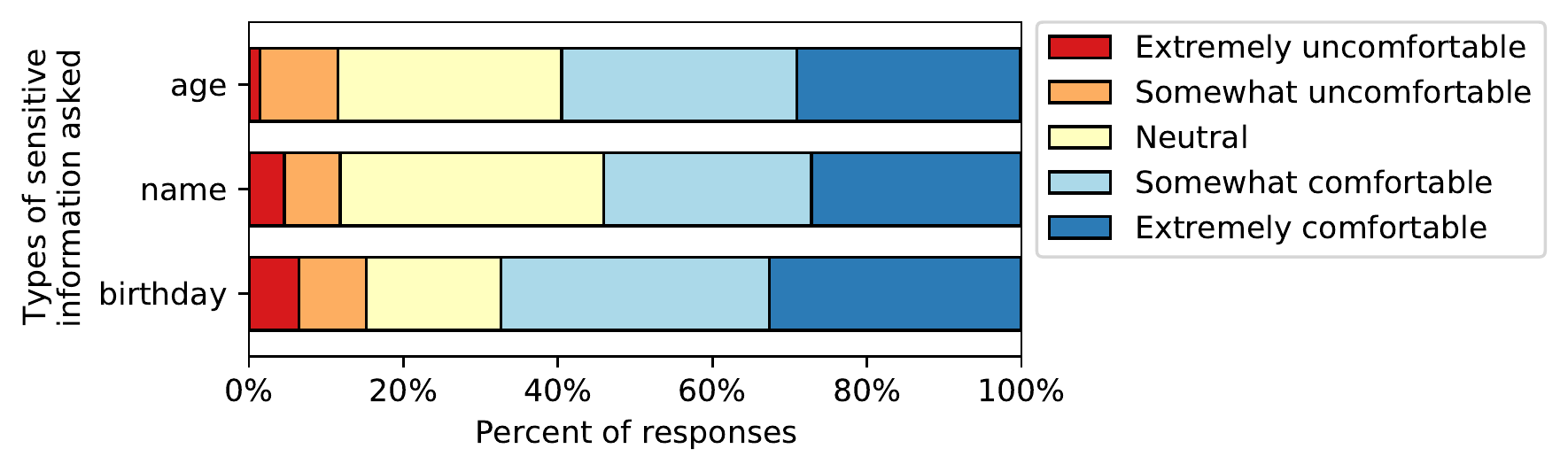}
  \caption{Participants' levels of comfort for each type of personal information, if the conversations happen between the participants' children and Alexa.}
  \label{fig:sensitive_labels}
\end{figure}

We do not find any significant difference between parents' comfort with the Sensitive conversations versus the Non-Risky conversations. However, the Sensitive conversations involved skills asking for different types of personal information. Out of the 20 skills in the Sensitive set, 15 skills asked for the user's name, 3 asked for the user's age, and 2 asked for the user's birthday. We show the distribution of the participants' comfort level according to each type of personal information in Figure~\ref{fig:sensitive_labels}. This indicates that that parents expressed more discomfort (``Extremely uncomfortable'' and ``Somewhat uncomfortable'') for skills that ask for the user's birthday (15.2\% of respondents), compared with skills that ask for the user's name (11.8\%) or age (11.5\%). Some participants expressed their concerns about these skills by free-text responses, including ``I don't like a skill or Alexa asking for PII (P115),'' ``I haven't had a similar experience but I think it is inappropriate for Alexa to be asking for the name of a child (P209),'' ``I don’t know why it needs a name (P228),'' and ``I would not want Alexa to collect my children's imformation [sic] (P003).''

\paragraph{\textbf{Amazon Echo Usage.}}
Our results also show that most households with children use Echo devices other than the Echo Kids Edition.
Echo Dot was the most popular type (46.4\%) of Echo device in our participants' households. Only 27 participants (6.8\%) bought an Echo Dot Kids Edition, which has parental control mode enabled by default. This shows that if children use an Echo, they likely have access to the types of Echo devices that do not have parental control mode enabled by default.

\begin{figure}[htbp]
  \centering
  \includegraphics[width=\columnwidth]{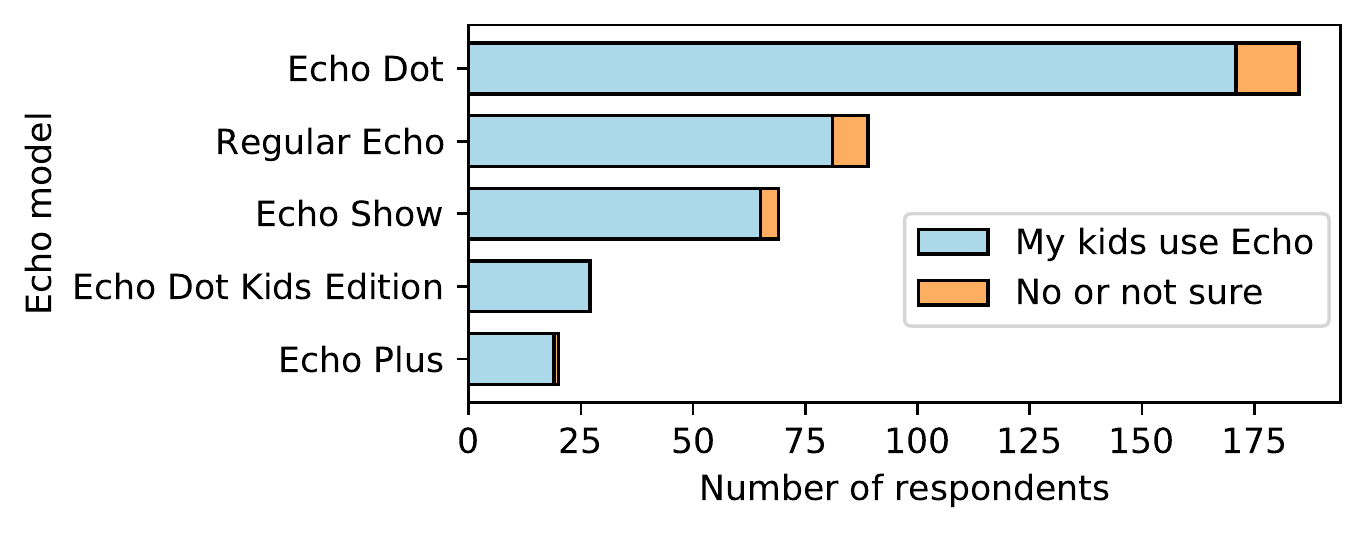}
  \caption{Types of Echo devices that the participants own in their households and the number of participants whose children use Amazon Echo at home. Echo Dot is the most popular device, and most households have children who use an Amazon Echo other than the Kids Edition.}
  \label{fig:echousage}
\end{figure}

Furthermore, the majority of participants (91.8\%) reported that their children do use Amazon Echo at home. 
Figure~\ref{fig:echousage} shows the types of Echo that the participants own in their household associated with the breakdown of answers to the question ``Do your kids use Amazon Echo at home?'' Most parents allow their children to use Amazon Echo at home even without an Echo Dot Kids Edition. This indicates that many children have access to risky skills, as these skills can be used by default on Echo devices other than the Kids Edition.

\paragraph{\textbf{Awareness of Parental Control Features.}}
We analyzed the responses to the question: ``Does Amazon Echo support parental control?'' In total, 76.3\% participants said ``yes,'' 0.4\% said ``no,'' and 23.3\% were unsure. For participants who had an Echo Kids Edition, almost all (92.6\%) said ``yes,'' 7.4\% said ``no,'' and none was unsure. In contrast, for participants without an Echo Kids Edition, only 74.1\% said ``yes,'' and 25.4\% were unsure. This indicates parents who did not buy the Echo Kids Edition are less likely to know about the parental control feature.

For those who said ``yes'' (i.e., they knew Amazon Echo supports parental control), we further asked if they used parental control. In total, only 29.4\% used parental control. Specifically, 64.0\% of participants that had Echo Kids Edition said they used parental control, but only 23.7\% of those who did not have Echo Kids Edition used parental control. Given that many fewer participants had an Echo Kids Edition (only 27 people out of 232), the majority of parents did not use parental control for their Echo devices at home. This result again indicates that children are more likely to have access to risky skills.

Furthermore, although some participants reported that they used parental control, many of them did not really know what the feature involved. For participants that said they used parental control, we asked them to tell us the name of the parental control feature. As long as their answer contained ``free'' and ``time'' (the correct answer is ``FreeTime''), we considered their answer correct. 66.7\% of participants that had an Echo Kids Edition were able to correctly name the parental control feature, but only 42.3\% of those that did not have an Echo Kids Edition could do so.

\paragraph{\textbf{Takeaway.}} Our results show that parents express significantly more disbelief and concern about conversations from Expletive skills versus Sensitive skills. This is worrisome, because all skills displayed on the survey are
real and can be invoked by \textit{anyone} now, including children. The fact that parents' responses to Sensitive skills were not significantly different from their responses to Non-Risky skills also illustrates a potential lack of understanding that skills are developed by third parties and may pose a privacy risk.
\section{Confounding Utterances Analysis}
\label{section:utterances}
To understand how easy it is for kids to accidentally trigger skills (\textbf{RQ3}), we performed a systematic analysis to identify confounding utterances and corresponding skills. We identified the set of potential confounding utterances from the skills' info pages, each of which might invoke multiple different skills, and used \systemname to collect and analyze conversations started by these utterances.

\subsection{Discovering Confounding Utterances}
To identify confounding utterances, we first created a dictionary with utterances extracted from Alexa skill info pages as keys and lists of skills that respond to each utterance as values. For each utterance, we removed punctuation and used lowercase to keep the format consistent. We filtered the dictionary to contain only the confounding utterances corresponding to at least two different skills. 
We then wanted to discover which skill Alexa chooses to enable/invoke when given a confounding utterance that is shared between multiple skills. When Alexa prioritizes a skill which does not match the user's intent, it poses a potential risk to the user. This is especially true if the user is a child. For example, if Alexa prioritizes a non-kid skill over a kid skill that shares the same utterance, a child user could inadvertently gain access to risky content.

We discovered a set of 4,487 confounding utterances, each of which was shared between two or more skills. Of these 4,487 utterances, 110 (2.5\%) belonged only to kid skills, 581 (12.9\%) belonged to both kid and non-kid skills, and 3,796 (84.6\%) belonged only to non-kid skills. We defined these three categories of confounding utterances as ``Kids Only,'' ``Joint,'' and ``Non-kids Only,'' respectively. We further identified utterances belonging to skills with the same skill name and skill icon---properties that would make it difficult for users to distinguish among these skills even by visually inspecting the skills' webpages. For confounding utterances in the Kids Only category, we found that 6 (5.5\%) out of 110 were shared among skills with the same name and icon. For those in the Joint category, we found that 48 (8.3\%) out of 581 were shared among skills with the same name and icon. For those in the Non-kids Only category, we found that 577 (15.2\%) out of 3,796 were shared among skills with the same name and icon.

\subsection{Testing Confounding Utterances}
We used \systemname to test the confounding utterances identified in the discovery step. For each utterance, we disabled all skills and entered the utterance. We then checked if any of the skills that shared the utterance had been enabled (Table~\ref{tab:confounding}).

\begin{table}[htbp]
\centering
\begin{tabular}{|l|c|c|c|c|}
\hline
\multicolumn{1}{|c|}{} & \makecell{Kids Only\\ (N=110)} & \makecell{Joint\\ (N=581)} & \makecell{Non-kids Only\\ (N=3,796)} & Total \\ \hline
\makecell{Invoked irrelevant skill} & 64 & 367 & 1,999 & 2,430 \\ \hline
\makecell{Invoked relevant skill} & 46 & 57 & 1,797 & 1,900 \\ \hline
\makecell{Invoked relevant skill\\ but prioritized non-kid skill} & - & 157 & - & 157 \\ \hline
\end{tabular}%
\caption{Number of confounding utterances of each type and corresponding behaviors.}
\label{tab:confounding}
\end{table}
\textit{\underline{Kids Only:}}
We found that 64 (58.2\%) out of the 110 confounding utterances in this category invoked an irrelevant skill that was not in the list of skills associated with the utterance itself. The remaining 46 utterances (41.8\%) invoked a relevant skill within the list of associated skills. Examples include the utterances ``Start owl facts'' (shared by two different kid skills) and ``Open animal sound quiz'' (shared by three different kid skills).

\textit{\underline{Both Kids and Non-kids (Joint):}}
We found that 367 (63.2\%) out of the 581 confounding utterances in this category invoked an irrelevant skill that was not in the list of skills associated with the utterance itself. The remaining 214 utterances (36.8\%) invoked a relevant skill within the list of associated skills. However, 157 out of these 214 utterances (73.4\%) prioritized invoking a non-kid skill over a kid skill. For example, ``Start human body quiz'' is a confounding utterance that is shared between two skills. One of them is a skill published for kids. The other one is a general game skill which is marked as having mature content.

\textit{\underline{Non-kids Only:}}
We found that 1,999 (52.7\%) out of the 3,796 confounding utterances in this category invoked an irrelevant skill that was not in the list of skills associated with the utterance itself. The remaining 1,797 utterances (47.3\%) invoked a relevant skill within the list of associated skills. Examples include ``Start movie picker'' (shared by five different skills) and ``Give me a random fact'' (shared by seven different skills).

\paragraph{\textbf{Takeaway:}}
It is risky if a confounding utterance is shared between a kid skill and a non-kid skill. Our analysis shows that children can accidentally invoke a non-kid skill while trying to use a kid skill. An adversary can exploit this problem to get child users to invoke risky non-kid skills.

\subsection{Confounding Utterances' Risks to Kids}
Our confounding utterance analysis shows that children can accidentally invoke a non-kid skill while trying to use a kid skill. Furthermore, confounding utterances could be an attack vector for an adversary to expose children to risky content without having to publish a kid skill, since 
non-kid skills do not need to follow the content requirements in place for kid skills. We discuss two risks that we observed during our confounding utterance analysis as follows.  

\paragraph{\textbf{Non-kid skills risks}} To gain more insights into the potential risks from non-kid skills, we sampled a set of non-kid skills containing 50 skills from each of the other 22 categories (1,100 skills in total). We ran \systemname to interact with each skill and record the conversations. We collected 7,356 conversations from 1,073 non-kid skills after removing 27 skills that resulted in errors or crashes. We used the same approach as our kid skill analysis presented in Section~\ref{analysis:riskykidskills} to analyze this sample. As a result, we identified 4 skills with inappropriate content and 16 skills asking for personal information, such as name, age, birthday, phone number, zipcode, and address. Of the 16 skills asking for personal information, 8 skills did not request any permissions from the user. We further investigated the 3 skills that requested permissions. Only one of them requested the proper permission for the personal information it queried.

\paragraph{\textbf{Sneaky skills risks}}
\label{sneakyskills}
During our analysis, we noticed an abnormal behavior of the Alexa skills interface. In particular, through the ``Your skills" tab on both Alexa web interface and mobile app, users can view a list of skills that are enabled in their account. We observed that the interface sometimes showed an empty list of skills although we actually had some skills enabled. We managed to further investigate this issue by enabling/disabling the skills via API calls and asking Alexa about the status of the skills to verify. We found that the skills were actually enabled and disabled properly in the backend, but the front-end did not display their status properly. From our user study, some participants mentioned in free-text responses that they preferred to monitor enabled skills on their own. Additionally, our skill analysis found that in some cases users have limited or no ability to check which skill they actually invoked due to skills having similar names and icons. Therefore, this front-end bug could become a security issue as it increases the stealthiness of malicious skills. In particular, a malicious skill can get accidentally enabled and remain invisible to users. In one possible attack, an adversary could craft a malicious skill that exploits a confounding utterance to become unexpectedly enabled/invoked by the user, causing the skill to become ``sneaky,'' as the front-end bug prevents the user from viewing what skills are enabled.
\section{Discussion}
\label{section:discussion}
This section provides some suggestions for building safe VPA for children. We also acknowledge some limitations of our study and discuss the potential future work.

\paragraph{\textbf{Suggestions for Building Safe VPA for Children.}}
Our study shows that the current Alexa skill vetting process is not effective, even for skills targeting children -- a vulnerable population. Although there are stricter requirements for developing and publishing kid skills, risky kid skills still exist on the store. This means VPA service providers need a more robust vetting system to ensure that the published skills adhere to policy requirements. Furthermore, since skills can be hosted on third-party servers, it is hard for VPA service providers to control what happens in the backend. A nefarious skill developer could easily manipulate the backend to turn a benign skill malicious. Thus, a continuous vetting process is important to ensure consistent adherence to policy requirements. 

VPA service providers also need to improve detection and limitation of confounding utterances, especially those that may unintentionally invoke different skills than the user intended. 
Without further protection, the existence of confounding utterances enables children to accidentally invoke non-kid skills, potentially exposing them to inappropriate content. 
Third-party skill designers could be required to register invocation phrases upon posting their skills on VPA provider hosting platforms. Like email addresses or domain names, preventing overlap of skill invocation phrases would keep children and other users from accidentally opening an unwanted skill with potentially risky content.

Our user study results showed that many parents do not know about or do not use parental control features on VPA devices. This likely means that more user-friendly parental control features are needed to reduce burden on parents while providing strong protections for child users. Parents should be encouraged to use parental control features for VPAs, especially on devices placed in a shared space. Existing recommendations for the design of parental control software in other domains likely carry over to the VPA space and would help parents' prevent children from accessing risky non-kid skills.

\paragraph{\textbf{Limitations.}}
The Alexa developer console employed by Skillbot has some limitations~\cite{alexasimulatorlimit} as compared to physical Echo devices. The limitation most relevant to this paper is the inability to collect the content of audio playbacks from skills (e.g., music skills). Thus, we did not consider such skills in our dataset. Since audio playbacks are rare (as shown in Section~\ref{system:evaluation:simulator}), we believe that this limitation is a reasonable trade-off for Skillbot's ability to efficiently perform analysis at scale and does not undermine our results. 

Our user study's results were based on self-reported data, which means the responses might be influenced by social desirability. Specifically, people are often biased in self-reports toward social norms~\cite{fisher2000social}. To mitigate this, we tried to use neutral wording for our questions. As with any self-reported surveys, participants may choose the first answer that satisfies without thinking carefully about the question.
Thus, we included attention check questions to filter out such inattentive participants from our study. In addition, our user study protocol includes incentives for survey completion, which might cause a bias. We also reviewed the survey responses and filtered out invalid responses as described in Section~\ref{userstudy:response-filtering}. We believe that if there are any remaining data quality issues, they are minor and do not affect our findings.

\paragraph{\textbf{Future Work.}}
We are willing to work with Amazon to resolve the issues presented in this paper. We also maintain a dataset of recorded conversations with Alexa skills for future research. This dataset will increase in size over time as we run Skillbot on new skills and longitudinally to explore new conversation branches of existing skills.
Future research using this dataset could provide insights into new and ongoing risky skill behaviors and help with the creation of rules and systems to protect users.

While this study focuses on risky content in kid skills, we show that children can potentially be exposed to non-kid skills through confounding utterances and a lack of parental controls. 
Thus, future work could further investigate the risks posed by non-kid skills that can be accidentally invoked by children.

Finally, we focused our analysis on Alexa, but it would be easy to extend Skillbot to work with other VPA platforms in order to investigate risky skills on those platforms. Skillbot could also be integrated into a public service for consumers to vet ``black-box'' skills before installing them (i.e., a potential countermeasure for consumers). 
\section{Related Work}
\label{section:relatedwork}
We discuss related work in this area and compare this prior research to our project. The related work can be categorized into three main streams as follows.

\subsection{Assessing Personal Assistants and Voice Apps}
Previous work looked into the problem of speech recognition misinterpretations made by voice personal assistants~\cite{kumar2019emerging, kumar2018squatting, zhang2019dangerousskills}, showing that an adversary could impersonate the voice assistant system or other skills to eavesdrop on users. Edu et al.~\cite{edu2021skillvet} showed that many developers implemented permission requests with bad privacy practices. Other work reversed engineered the Alexa skill vetting process and revealed limitations that allowed policy-violating skills to be published~\cite{lentzsch2021hey, cheng2020certified}. Lentzsch et al.~\cite{lentzsch2021hey} and Liao et al.~\cite{liao2020measuring} also found that many skills did not provide a valid privacy policy. Unlike these previous papers, we investigate the hidden logic inside skills to identify risky content. Recently, Guo et al.~\cite{guo2020skillexplorer} investigated skills asking users' for private information by triggering the skills starting with the three sample utterances provided by the skills themselves. 

Our work focuses on children's safety when using voice apps, including a broader set of risks than explored by~\cite{guo2020skillexplorer}, including inappropriate content and asking for personal information. Our \systemname system also explores the possible conversations that users can have with Alexa skills in much greater depth and breadth via our chatbot module. This is critical for identifying risky skill behaviors, especially in kid skills that have to follow stricter policies to get published. For example, out of the 28 risky kid skills (8 expletive and 20 sensitive) identified in our study, 3 required the use of custom utterances generated by Skillbot. In addition, 2 risky skills were only identified at a depth of 11 in the conversation tree, 1 skill at depth 5, 4 skills at depth 4, 6 at depth 3, 8 at depth 2, and 7 at depth 1. Skillbot's ability to conduct such extended back-and-forth interactions with skills separates it from prior work in the area. 
We also ran a user study to understand parents' experiences, awareness, concerns about risky content for children, and use of parental control. In addition, we described and analyzed confounding utterances---a novel threat which exposes children to risky contents. 

\subsection{Perceptions of Personal Assistants}
Major et al.~\cite{major2019alexa} found that Alexa users often confused third-party skills with built-in Alexa features, and they did not know what features the native Alexa system supports. Abdi et al.~\cite{abdi2019more} found that users have different perceptions of how data is processed by personal assistants and that users have security/privacy concerns about what the assistants learn. They further explored the privacy norms for personal assistants in smart homes and visualized the acceptability of information flows based on contextual integrity~\cite{abdi2021privacy}. Pins et al.~\cite{pins2021alexa} designed an interactive approach to visualize conversations with voice assistants and help users with data literacy. Meng et al.~\cite{meng2021owning} investigated the perceptions of ownership for personal assistants in multi-user smart homes and showed that most users felt like an owner of the shared device although they did not have the same rights or controls. Lau et al.~\cite{lau2018alexa} found non-users did not trust the smart speaker companies while users expressed few concerns but did not completely understand the risks. They also highlighted that users often trade privacy for convenience and the current privacy controls are not effective. Different from these prior works, our user study focuses on the impacts of risky skills on children. We find that children often use types of Echo devices other than the Kids Edition in their households and that parents have concerns about the risky skills we identified. Additionally, many parents do not think such risky skills are available to children although those skills are actually published for children.

\subsection{Children and the Internet}
There is considerable literature examining children's online protection and COPPA compliance in domains other than VPA. Automated analyses of thousands of mobile applications have found widespread COPPA violations~\cite{zimmeck2016automated, reyes2018won} ranging from illegal collection of persistent identifiers to non-compliant privacy policies. Investigations of children-directed websites have also found covert tracking techniques designed to avoid COPPA requirements~\cite{vlajic2018online}, among other non-compliant behaviors~\cite{turow2001privacy, cai2013online}. Our work sheds light on the new problems in VPA. 
More recently, researchers have raised concerns about children's privacy with respect to Internet-connected ``smart'' toys. Several studies have conducted detailed analyses of specific toys, often noting multiple implementation decisions and security vulnerabilities that place children's data at risk and violate COPPA~\cite{chu2018security, mahmoud2018towards, streiff2018s, valente2017security}. Other studies have provided frameworks for smart toy protections~\cite{rafferty2017towards, haynes2017framework}, and recommendations for smart toy manufacturers. As VPAs straddle the boundary between IoT products, mobile application platforms, and search engines, our work contributes to the evolving landscape of risks posed to children by modern online services, joining this related literature to demonstrate the breadth of challenges remaining. 

Importantly, technical research into children and the Internet has been motivated and supported by qualitative and quantitative user studies investigating how children and parents understand connected technologies~\cite{mertala2019young} and make privacy decisions. These studies show that parents are becoming more concerned about smart toy privacy~\cite{manches2015three, mcreynolds2017toys}, but that children have difficulty conceptualizing certain types of privacy risks~\cite{Zhao:2019:IMU:3290605.3300336, kumar2017no}. Given that some parents actively compromise their children's online privacy~\cite{minkus2015children} or help their children avoid COPPA protections~\cite{hargittai2011parents}, we should not rely on parents to keep children safe online. Instead, risky practices, such as the skills we identify, must be addressed through a combination of academic, regulatory, and industry action.
\section{Conclusion}
\label{section:conclusion}
We designed and implemented an automated skill interaction system called \systemname, analyzing 3,434 Alexa kid skills. We identified a number of risky skills with inappropriate content or personal data requests and described a novel confounding utterance threat. To further evaluate the impacts of these risky skills on children, we conducted a user study of 232 U.S. parents who use Alexa in their households. We found widespread concerns about the contents of these skills, combined with general disbelief that these skills might actually be available to children and a low rate of adoption of parental control features.

\begin{acks}
This work was supported in part by the National Science Foundation Grants NSF 1943100, NSF 1920462, NSF 2114074, and the Facebook Faculty Fellowship. Any findings, conclusions, or recommendations presented in this paper are those of the authors, and do not necessarily reflect those of the funding agencies.
\end{acks}

\bibliographystyle{ACM-Reference-Format}
\bibliography{toit-main}

\appendix

\section{Survey Questionnaire}
\label{appendix:questionaire}

\subsection{Screening Survey}

\Qitem{ \Qq{Who lives in your household? (Choose all that apply)}
\begin{Qlist}
\item Myself
\item My spouse or partner
\item My friend(s)
\item My sibling(s)
\item My kid(s) - aged 1 to 13
\item My kid(s) - aged 14 to 18
\item My parent(s)
\item My grandparent(s)
\item Housemate(s) or roommate(s)
\item Other relative(s)
\item Other non-relative(s)
\end{Qlist}
}

\Qitem{ \Qq{Which type of electronic devices do you have in the household? (Choose all that apply)}
\begin{Qlist}
\item Amazon Echo
\item Google Home
\item Smart TV
\item Computer
\item Smartphone
\item Other: \Qline{4cm}
\item None of the above
\end{Qlist}
}

\subsection{Main Survey}
\setcounter{itemnummer}{0}

\subsubsection{Parents' Reactions to Risky Skills}
For each participant, show the following skills in random order and present the set of questions below.
\begin{itemize}
    \item Skill 1: Randomly selected from non-risky set
    \item Skill 2: Randomly selected from non-risky set
    \item Skill 3: Randomly selected from sensitive set
    \item Skill 4: Randomly selected from sensitive set
    \item Skill 5: Randomly selected from expletive set
    \item Skill 6: Randomly selected from expletive set
\end{itemize}

\Qitem{ \Qq{Do you think this conversation is possible on Alexa?}
\begin{Qlist}
\item Yes
\item No
\item Not sure
\end{Qlist}
}

\Qitem{ \Qq{Do you think Alexa should allow this type of conversation?}
\begin{Qlist}
\item Yes
\item No
\item Not sure
\end{Qlist}
}

\Qitem{ \Qq{Do you think this particular skill or conversation is designed for families and kids?}
\begin{Qlist}
\item Yes
\item No
\item Not sure
\end{Qlist}
}

\Qitem{ \Qq{How comfortable are you if this conversation is between your children and Alexa?}
\begin{Qlist}
\item Extremely uncomfortable
\item Somewhat uncomfortable
\item Neutral
\item Somewhat comfortable
\item Extremely comfortable
\end{Qlist}
}

If answering ``Somewhat uncomfortable" or ``Extremely uncomfortable", ask:
\Qitem{ \Qq{What skills or conversations have you experienced with Alexa that made you similarly uncomfortable?}
\Qline{4cm}
}

\subsubsection{Amazon Echo Usage}

\Qitem{ \Qq{Which model(s) of Amazon Echo do you have in the household? (Choose all that apply)}
\begin{Qlist}
\item Regular Echo
\item Echo Dot
\item Echo Dot Kids Edition
\item Echo Plus
\item Other: \Qline{4cm}
\end{Qlist}
}

\Qitem{ \Qq{Do your kids use Amazon Echo at home?}
\begin{Qlist}
\item Yes
\item No
\item I don't know
\end{Qlist}
}

\subsubsection{Awareness of Parental Control Features}

\Qitem{ \Qq{Does Amazon Echo support parental control?}
\begin{Qlist}
\item Yes
\item No
\item I don't know
\end{Qlist}
}

If Yes, ask:
\Qitem{ \Qq{Do you use Amazon Echo's parental control?}
\begin{Qlist}
\item Yes
\item No
\item I don't know
\end{Qlist}
}

If Yes, ask:
\Qitem{ \Qq{What is the name of Amazon Echo's parental control?}
\Qline{4cm}
}

\subsubsection{Demographic Information}

\Qitem{ \Qq{What is your gender?}
\begin{Qlist}
\item Male
\item Female
\item Other: \Qline{4cm}
\item Prefer not to answer
\end{Qlist}
}

\Qitem{ \Qq{What is your age?}
\begin{Qlist}
\item 18 - 24 years old
\item 25 - 34 years old
\item 35 - 44 years old
\item 45 - 54 years old
\item 55 - 64 years old
\item 65 - 74 years old
\item 75 years or older
\end{Qlist}
}

\Qitem{ \Qq{Please select the statement that best describes your comfort level with computing technology.}
\begin{Qlist}
\item Ultra Nerd: I build my own computers, run my own servers, code my own apps. I’m basically Mr. Robot.
\item Technically Savvy: I know my way around a computer pretty well. When anyone in my family needs technical help, I’m the one they call.
\item Average User: I know enough to get by.
\item Luddite: Technology scares me! I only use it when I have to.
\end{Qlist}
}


\section{Examples of Skills in the Expletives and Sensitive Sets}
\label{skillexamples}

\subsection{Expletive Set}

Skill ID: B07CX9G446 -- You: ``open word rally'' -- Alexa: ``Poop.''

Skill ID: B07K77Z4F8 -- You: ``open fart buttons'' -- Alexa: ``Welcome to the Fart Buttons Skill. Press your Echo Buttons to play a fart sound..''

Skill ID: B07RYKZW52 -- You: ``open blast off'' -- Alexa: ``Ok, Here's Flaming Fart.''

Skill ID: B07ZCS7784 -- You: ``open new facts'' -- Alexa: ``Here's your fact: A pig's orgasm lasts for 30 minutes..''

Skill ID: B081VS1BTD -- You: ``play santa claus a fart'' -- Alexa: ``Handing off to SantaClausAFart by Baomatic Apps.''

Skill ID: B0837HWNY5 -- You: ``open my burns'' -- Alexa: ``You're so ugly you'd scare the crap out of the toilet. I'm on a roll.''

Skill ID: B0881V9K8V -- You: ``open birthday countdown'' -- Alexa: ``... until November 5! In that time, you will fart approximately 3 thousand times. Do you want me ...''

Skill ID: B08C4WVY4Q -- You: ``start pop story'' -- Alexa: ``... story, I will read it to you now. poop a poop Once in a faraway kingdom there lived a Poop and Poopy who were the shortest in the land, and ... and humongous pizzas. aw manWell, one day, the poop and poopy had a wonderful daughter, who they named a ... by the time she was a teenager. The ...''   
    
\subsection{Sensitive Set}

Skill ID: B0773H1M6J -- You: ``start spelling bee'' -- Alexa: ``... understand what you said, please tell me your name. for example, I am Cindy..''

Skill ID: B077C9B1KX -- You: ``open spelling contest'' -- Alexa: ``... say ''start'' to start a contest, or my name is to change your name.''

Skill ID: B078XHP4MQ -- You: ``open does it fly'' -- Alexa: ``... play does it fly game, tell me your name.. So more, Alexa will ask you a question,''

Skill ID: B078XQC43F -- You: ``open mr. tongue twister'' -- Alexa: ``Welcome to Mr Tongue Twister. What is your name?.''

Skill ID: B079CMS36Z -- You: ``ask find buddy to look in the garden'' -- Alexa: ``... - if you say ''call me'' before your name, I may understand you better. For example, if your name is Allen, try saying ''Call me Allen'', instead''

Skill ID: B07CQ68G5C -- You: ``open ready freddy'' -- Alexa: ``... Ready Freddy! Please introduce yourself by saying: My name is, and then your name..''

Skill ID: B07CQQWWRT -- You: ``open birthday wisher'' -- Alexa: ``Hello! I am Birthday Wisher , by Uday Pratap Singh. What is your birthday?.''

Skill ID: B07DB9JMCP -- You: ``open community helpers'' -- Alexa: ``... I didn't get that. Would you state your name again?.''

Skill ID: B07DFCXXM5 -- You: ``open say please'' -- Alexa: ``Welcome to Say Please! My name is Sara. Please tell me your name..''

Skill ID: B07L9QRYNH -- You: ``open santa's little helper'' -- Alexa: ``... to help you with that. Tell me your name to see if you are on the naughty''

Skill ID: B07RZS3PGV -- You: ``start may twenty question'' -- Alexa: ``You said your name is 'John'. Is that correct?.''

Skill ID: B07YYK45WQ -- You: ``start my birthday saifi'' -- Alexa: ``You said your name is 'John'. Is that correct?.''

Skill ID: B0819YKNHJ -- You: ``open push up'' -- Alexa: ``... the Push Ups Counter. May I have your name, please?.''

Skill ID: B087J97GBX -- You: ``ask banana stories to start the game'' -- Alexa: ``So, first, what is your name?.''

Skill ID: B088TQHVCJ -- You: ``open birthday surprise'' -- Alexa: ``Hello! I'll remember your birthday when no one else will! This is Birthday Surprise, when is your birthday?.''

Skill ID: B08C7CC4CC -- You: ``open your name says'' -- Alexa: ``Welcome to Your Name Says!Let's start by you saying your name...What's your name?.''

Skill ID: B08DD9LSB4 -- You: ``open hamilton knowledge'' -- Alexa: ``I have heard your name before John.''

Skill ID: B07MJZ5R65 -- You: ``open my book assistant'' -- Alexa: ``Please provide me your age group. Such as adult, children, young Et cetera..''

Skill ID: B07TZM51S4 -- You: ``open shape game'' -- Alexa: ``Awesome! Before we start however; I'm curious...how old are you?.''

Skill ID: B07XV6Z4ZM -- You: ``open guess my age'' -- Alexa: ``First you must write down your age. Say the word READY to continue..''

\section{Confusion Matrix for Skill Response Classification}
\label{appendix:classificationmatrix}
\begin{table}[htbp]
\centering
\small
\resizebox{\linewidth}{!}{%
\begin{tabular}{lcccc}
\cline{1-5}
\multicolumn{1}{|c|}{} & \multicolumn{1}{c|}{Actual Yes/No} & \multicolumn{1}{c|}{Actual Open-ended} & \multicolumn{1}{c|}{Actual Non-question} & \multicolumn{1}{c|}{Total Predicted} \\ \cline{1-5}
\multicolumn{1}{|l|}{Predicted Yes/No} & \multicolumn{1}{c|}{51} & \multicolumn{1}{c|}{0} & \multicolumn{1}{c|}{5} & \multicolumn{1}{c|}{56} \\ \cline{1-5}
\multicolumn{1}{|l|}{Predicted Open-ended} & \multicolumn{1}{c|}{0} & \multicolumn{1}{c|}{47} & \multicolumn{1}{c|}{3} & \multicolumn{1}{c|}{50} \\ \cline{1-5}
\multicolumn{1}{|l|}{Predicted Non-question} & \multicolumn{1}{c|}{1} & \multicolumn{1}{c|}{3} & \multicolumn{1}{c|}{190} & \multicolumn{1}{c|}{194} \\ \cline{1-5}
\multicolumn{1}{|c|}{Total Actual} & \multicolumn{1}{c|}{52} & \multicolumn{1}{c|}{50} & \multicolumn{1}{c|}{198} & \multicolumn{1}{l|}{} \\ \cline{1-5}
\end{tabular}%
}
\caption{Skill response classification confusion matrix}
\label{tab:classificationmatrix}
\end{table}

\section{Confusion Matrix for Identifying Risky Skills Asking for Personal Information}
\label{appendix:riskyidentificationmatrix}
\begin{table}[htbp]
\centering
\begin{tabular}{lllll}
\cline{1-3}
\multicolumn{1}{|l|}{}                    & \multicolumn{1}{l|}{Actual risky} & \multicolumn{1}{l|}{Actual non-risky} &  &  \\ \cline{1-3}
\multicolumn{1}{|l|}{Predicted risky}     & \multicolumn{1}{l|}{20}           & \multicolumn{1}{l|}{2}                &  &  \\ \cline{1-3}
\multicolumn{1}{|l|}{Predicted non-risky} & \multicolumn{1}{l|}{0}            & \multicolumn{1}{l|}{100}              &  &  \\ \cline{1-3}
                                          &                                   &                                       &  & 
\end{tabular}
\caption{Confusion matrix for identifying risky skills that ask for personal information. 20 skills were correctly predicted as risky while 2 skills were false positives. There were 0 false negatives, as all 100 non-risky skills were correctly predicted.}
\label{tab:risky-personal-info-confusion-matrix}
\end{table}

\end{document}